\documentclass[12pt,preprint]{aastex}

\usepackage{natbib}
\usepackage{graphics,epsfig}
\usepackage{multirow}
\usepackage{amsmath}
\usepackage{xcolor}

\newcommand{\degree}{\ensuremath{^\circ}}

\def\spose#1{\hbox to 0pt{#1\hss}}
\def\lta{\mathrel{\spose{\lower 3pt\hbox{$\mathchar"218$}}
     \raise 2.0pt\hbox{$\mathchar"13C$}}}
\def\gta{\mathrel{\spose{\lower 3pt\hbox{$\mathchar"218$}}
     \raise 2.0pt\hbox{$\mathchar"13E$}}}

\shorttitle{Detection of Exomoons}
\shortauthors{Noyola et al.}

\begin{document}
\title{Detection of Exomoons Through Observation of Radio Emissions}

\author{J. P. Noyola, S. Satyal and Z. E. Musielak}
\affil{The Department of Physics, University of Texas at Arlington, 
       Arlington, TX 76019;\\ 
       joaquin.noyola@mavs.uta.edu; ssatyal@uta.edu; zmusielak@uta.edu}

\begin{abstract}
In the Jupiter-Io system, the moon's motion produces currents along the field lines that connect it to Jupiter's polar regions. 
The currents generate, and modulate radio emissions along their paths via the electron-cyclotron maser instability. 
Based on this process, we suggest that such modulation of planetary radio emissions may reveal the presence of exomoons around giant planets in exoplanetary systems. 
A model explaining the modulation mechanism in the Jupiter-Io system is extrapolated, and used to define criteria for exomoon detectability. 
A cautiously optimistic scenario of possible detection of such exomoons around Epsilon Eridani b, and Gliese 876 b is provided.  
\end{abstract} 

\keywords{Method: analytical ---
Exomoons: detection}

\section{Introduction}
Since the discovery of the first confirmed extra solar planet (exoplanet) around the pulsar PSR1257 + 12 by \cite{wol92}, there has been great progress in detection techniques and instrumentation, resulting in hundreds of confirmed exoplanets\footnote{http://exoplanet.eu/catalog/}, and thousands of exoplanet candidates identified by the NASA's {\it Kepler}\footnote{http://kepler.nasa.gov/Mission/discoveries/candidates/} space telescope.
Nonetheless, the current limits of observational techniques have not made it possible to confirm any exomoon detection.

Based on our knowledge of the Solar System, one might expect exomoons to be present around some of the known exoplanets, and several potential candidates have already been suggested. 
For example, orbital stability criterion was used by \cite{qua12} and \cite{cun13} to suggest the existence of exomoons in the Kepler 16 and HD 23079 planetary systems, respectively. 
Moreover, \cite{kip09} suggested that exomoons may actually be discovered in the data already collected by the {\it Kepler} mission. 
Among the moons of our Solar System, there is an interesting example of a planet-moon interaction observed in the Jupiter-Io system, where Io's motion inside Jupiter's magnetosphere induces radio emissions \citep{big64}. 
The motion produces currents along the field lines that connect Io to the Jupiter's polar regions, where the radio emissions are modulated by said currents \citep{acu81, mau01}.

In this study, we use Io-controlled decametric emissions (Io-DAM) as a basis to demonstrate how the presence of exomoons around giant exoplanets may be revealed by the same modulation mechanism.  
We determine the required physical conditions for such interaction, then assess the feasibility of our model by providing a cautiously optimistic scenario whereby exomoons could be found. 
Then, we use the available information on the proposed Square Kilometer Array(SKA\footnote{http://www.skatelescope.org/}) radio telescope as an example of what kind of exomoons could be detected if such technologies were fully implemented.
Furthermore, we apply our results to the two nearby exoplanetary systems Epsilon Eridani, and Gliese 876 to show that finding exomoons in this systems is not beyond the realm of possibility.
Finally, we discuss future improvements to our model, and how those improvements might change the results presented here.

There are several previous studies that are at least partially based on the Jupiter-Io system, and in which the authors suggest detecting exoplanets by using exoplanetary radio emissions \citep[and references therein]{laz04}. 
However, those authors based their studies on the non-Io-controlled decametric (non-Io DAM) Jovian emissions, instead of the Io-DAM, which originate directly from interactions between Io and the Jovian magnetosphere. 
Moreover, \cite{nic11} studied the Jupiter’s magnetosphere-ionosphere coupling mechanism, and hypothesized that this mechanism could produce enough radio power to be detectable from Earth.
Both his study and ours are based on the unipolar inductor mechanism, and the current source is Io’s plasma torus.
However, the circuit made by the current in each case is fundamentally different.  In our study, the circuit directly couples Io to Jupiter’s poles, whereas in Nichols's study the current bypasses Io, and instead flows through magnetic field lines well beyond the moon.

As with the other studies mentioned above, Nichols's mechanism requires a large stellar luminosity.
Specifically, it requires large X-ray and EUV stellar irradiation of the exoplanet’s ionosphere to produce a large enough output power to make it detectable.
Our detection method, as it is explained in this paper, does not have such a requirement, and in fact favors low stellar irradiation scenarios.
More importantly, even though Nichols emphasizes active moons as the sources of plasma, this might not be the case.
Recent computational studies have shown that stellar irradiation alone can ionize the hydrogen-rich atmosphere of a Jovian exoplanet to levels that can match, and even vastly exceed, the amount of plasma in Io’s plasma torus \citep{tra14}.
In other words, Nichols's mechanism cannot be used to detect exomoons, but only exoplanets.

This paper is organized as follows. In Section 2 we present the basic theory on the Io-Jupiter system, intensity of radio emissions, and the magnetic fields in giant planets.
In Section 3 we present our results on detectability scenario of exomoons followed by discussion.
We conclude in Section 4 with a brief overview of our results.

\section{Theory}
\subsection{The Io-Jupiter System}
Io is an intensely volcanic moon orbiting inside Jupiter's magnetosphere.
The volcanic activity creates a light atmosphere of SO$_{2}$ around Io, which ionizes to create an ionosphere \citep{lop07}.
This ionosphere then injects ions into Jupiter's magnetosphere to create a plasma torus, which orbits Jupiter's magnetic equator at an angle of 9.6$^\circ$ from the rotational equator, and co-rotates with the magnetic field at a speed of 74 $kms^{-1}$ \citep{su09}.
Io orbits Jupiter at a linear speed of 17 $kms^{-1}$, so Jupiter's magnetic field passes Io at a speed of 57 $kms^{-1}$. 
The speed difference gives rise to a unipolar inductor \citep{gri07}, which induces a current across Io's atmosphere of a few million amps.
The current then accelerates the electrons that produce the characteristic radio emissions \citep{cra97}.
It must be noted that while volcanism is essential to the formation of a dense ionosphere around Io, such process might not be required for larger moons, since moons like Titan are already large enough to sustain a thick atmosphere, which in turn can give rise to an ionosphere. 

Furthermore, the interaction between Io and the plasma torus gives rise to Alfv\'en waves \citep{bel87}.
The precise mechanism by which Alfv\'en waves interact with the torus is complex, and several analytical and numerical models have been proposed.
In these models, Alfv\'en waves produce electric fields parallel to Jupiter's magnetic field lines, which then transport and accelerate electrons towards Jupiter's magnetic poles \citep[and references therein]{su09, sau99, cra97, neu80}.
The electrons traveling through the field lines create a cyclotron maser which then emits radio waves whose existence in the Jupiter-Io system has been observationally verified \citep{cra97, mau01}.

Our studies of exoplanet-moon interactions are based on an extrapolation of both Io's plasma environment, and Jupiter's magnetic field to different scenarios that could potentially be encountered in newly discovered planetary systems. 
We begin by finding an expression for the maximum intensity of the radio emissions, then proceed to address the magnetic field and plasma properties, and finally analyze the dependence of the radio emissions on these parameters.

\subsection{Intensity of Radio Emissions}

Assuming a simple current distribution around Io, and that the magnetic field lines are approximately perpendicular to the plasma velocity, \cite{neu80} found that the maximum Joule dissipation of the system is given by

\begin{equation}
P_{T}=\frac{\pi R_{Io}^{2}V_{0}^{2}B_{Io}^{2}}{\mu_{0}\sqrt{V_{A}^{2}+
V_{0}^{2}}}\ ,
\label{eq:P_T}
\end{equation}

\noindent
where $R_{Io}$ is Io's radius, $B_{Io}$ is the the magnetic field at Io's position, $V_{0}$ is the plasma speed relative to Io, $\mu_{0}$ is the permeability of free space, and $V_{A}$ is the Alfv\'en velocity.
The Alfv\'en velocity depends on the magnetic field, $B_{Io}$, and the plasma density, $\rho_{Io}$, through the relationship $V_{A}=B_{Io}/\sqrt{\mu_{0}\rho_{Io}}$.

Since only a small fraction of the maximum Joule dissipation, $P_{T}$, is converted to radio waves, we will introduce the efficiency coefficient $\beta_{Io}$ which, based on previous studies, is $\approx$1\% for the Jupiter-Io system \citep{zar01}.
There is little information on the variability of this parameter, so we assume that other exoplanet-moon systems have similar efficiency coefficients. 
Hence, the maximum radio emission intensity, $P_{rad}$, from these systems is given as

\begin{equation}
P_{rad}=\frac{\pi\beta_{S}R_{S}^{2}V_{0}^{2}B_{S}^{2}}{\mu_{0}\sqrt{B_{S}^{2}
/(\mu_{0}\rho_{S})+V_{0}^{2}}}\ ,
\label{eq:P_rad_general}
\end{equation}

\noindent
where the subscript "Io" was switched to "S" to denote that these variables now belong to a generic exomoon, or "satellite."
The plasma speed, $V_{0}$, is computed assuming it corotates with the planet's magnetic field, as in the Jupiter-Io case.
Explicitly, if the moon orbits at a distance $r_{S}$ from the planet, then $V_{0}=\omega_{P}r_{S}-\sqrt{GM_{P}/r_{S}}$, where $G$ is the gravitational constant, $M_{P}$ is the planet's mass, and $\omega_{P}$ is the planet's angular velocity. 
The other parameters in Eq. (\ref{eq:P_rad_general}) are explored in later sections.

Equation (\ref{eq:P_rad_general}) does not depend on the properties of the host star, and therefore the exoplanet-moon system does not have to be close to its star (or even have a host star) to be detectable. 
In fact, exomoons around exoplanets with small orbits might be undetectable because stellar winds can also induce radio emissions, which increase with decreasing planetary semi-major axis \citep{zar98}. 
Furthermore, moon-induced radio emissions do not solely occur along the orbital plane, which means the system's bodies do not have to orbit parallel to our line of sight.

The dependence of Eq. (\ref{eq:P_rad_general}) on $R_{S}^{2}$ clearly favors large exomoons.
Although there has been no observational evidence, the possibility of detecting such exomoons is still plausible.
Nevertheless, given that mass grows as the cube of the radius, the long-term stability of such systems can also be called into question; however, orbital stability analysis is beyond the scope of this study, and it will not be discussed any further.

\subsection{Magnetic Fields in Giant Planets}

All the giant planets and several rocky bodies in the Solar System have magnetic fields and extended magnetospheres, of which Jupiter's is the largest.
Consequently, one can expect exoplanets to have similar magnetic fields and extended magnetospheres.  
To model an exoplanet's magnetic field, we begin by assuming that the field is mostly dipolar, as is the case for all magnetized bodies in the Solar System, and that its angle of inclination with respect to the exoplanet's axis of rotation is small enough to be neglected.
If the exomoon orbits close to the rotational equator of the exoplanet, then the magnetic field affecting the exomoon at its location is given by $B_{S}$ = $(\mu_0/4\pi)(m/{r_S^3})$, where $m$ is the exoplanet's magnetic dipole moment.

To approximate the magnetic moment of an exoplanet, we adopt the approximation introduced by \cite{dur09} who found that the magnetic moment of a planetary body can be expressed as $m=10^{-5}(\sigma M_{P}/T_{P})^{K}$, where the exponent $K$ is experimentally determined to be 1.10 $\pm$ 0.13 (we use 1.15 to better fit the giant planets), $T_{P}$ is the planet's rotation period, and $\sigma$ is the conductivity of the liquid at its core which is responsible for creating the magnetic field .
In the case of giant planets, the liquid that creates their magnetic field is metallic hydrogen, which is estimated to have a conductivity of about 2 $\pm$ 0.5$\times 10^{5}$ S/m \citep{shv07}.
The expression for the magnetic field affecting the exomoon is thus
\begin{equation}
	B_{S}=\frac{10^{-12}}{r_{S}^{3}}\left(\frac{M_{P}\sigma}{T_{P}}\right)^{K}.
\label{eq:BS_final}
\end{equation}

It must be mentioned that there are other approximations that we could have chosen to calculate $m$, but we chose Durand-Manterola's formula because of its simplicity, and because it can be at least partially justified using standard electrodynamics.
If the exomoon is sufficiently large, it could also have its own magnetic field, but here we assume its effects to be negligible.
However, based on the interaction between planets and the solar wind, we hypothesize that the exomoon's magnetic field creates a bow shock where the exomoon's magnetic field pressure equals the plasma torus pressure, and its net effect is to increase the apparent cross sectional area of the exomoon, thereby increasing the power of its radio emissions.
This effect will be quantitatively discussed in later publications.

\subsection{Other Parameters and Assumptions}

The intensity of exomoon-induced radio emissions depends on many parameters, thus a thorough clarification of how each of these parameters are treated is crucial.
A very important parameter in our calculation of a radio signal's flux is $T_{P}$, but it is also very difficult to measure or predict. 
Therefore, it is typically assumed to be equal to Jupiter's rotational period, $T_{J}$, or that the planet is tidally locked if it is closer than 0.1 AU to its host star \citep[and references therein]{laz04}. 
Since we mostly consider exoplanets with large orbits, we assume $T_{P}=T_{J}$ throughout the calculations.

The atmospheric plasma density of an exomoon (or exoplanet) is difficult to determine if the environmental properties of the body are not already known.
Even though we cannot predict the plasma density of a hypothetical exomoon, we can find a reasonable estimate based on what we know about the Solar System.
The plasma density depends not only on the number of ions present, but also on the molecular weight of the ions that constitute it.
For example, on Io one can find O$^{+}$, S$^{+}$, SO$^{+}$, etc, because Io's atmosphere is made from the SO$_{2}$ emitted by its volcanoes \citep{su09}.
Io's mean plasma density is $\sim 4.2\times10^{4}$ $amu$ $cm^{-3}$, or $\sim 7\times10^{-17} kgm^{-3}$ \citep{kiv04}.
Earth's ion number density is typically on the order of $10^{5} cm^{-3}$, and the dominant ion is O$^{+}$ which gives us a plasma density of $\sim 1.7\times10^{6}$ $amu$ $cm^{-3}$, or $\sim 2.7\times10^{-15} kgm^{-3}$ \citep{sch09}.
Venus, and Mars also have similar plasma densities to Earth, with O$_{2}^{+}$, and O$^{+}$ ions from CO$_{2}$ being the dominant species in their ionospheres \citep{sch09}.
Given the large variability of this parameter, we chose to work with three different values: $10^{4} amu$ $cm^{-3}$, $10^{5} amu$ $cm^{-3}$, $10^{6} amu$ $cm^{-3}$, which cover most of the range of plasma densities that can lead to a detectable exomoon with reasonable size (as explained in the next section).
We can see the dependence of $P_{rad}$ on the plasma density, $\rho_{S}$, by rearranging Eq. (\ref{eq:P_rad_general}) to get

\begin{equation}
P_{rad}=\frac{\pi\beta_{S}R_{S}^{2}B_{S}^{2}V_{0}}{\mu_{0}}\sqrt{\frac{\rho_{S}}
{\rho_{S}+\frac{1}{\mu_{0}}\left(\frac{B_{S}}{V_{0}}\right)^{2}}}.
\label{eq:P_rad_rho}
\end{equation}

Equation (\ref{eq:P_rad_rho}) allows us to define a critical density $\rho_{C}=\mu_{0}^{-1}(B_{S}/V_{0})^{2}$, which in turn allows us to characterize the three limiting cases:

\begin{eqnarray}
P_{-} &\equiv& P_{rad}(\rho_{S} << \rho_{C}) = \pi\beta_{S}R_{S}^{2}V_{0}^{2}B_{S}\sqrt{\frac{\rho_{S}}{\mu_{0}}},\\
P_{C} &\equiv& P_{rad}(\rho_{S} = \rho_{C}) = \frac{1}{\sqrt{2}}\pi\mu_{0}^{-1}\beta_{S}R_{S}^{2}B_{S}^{2}V_{0},\\
P_{+} &\equiv& P_{rad}(\rho_{S} >> \rho_{C}) = \pi\mu_{0}^{-1}\beta_{S}R_{S}^{2}B_{S}^{2}V_{0}.
\label{eq:PC}
\end{eqnarray}

The fact that $P_{C} = \frac{1}{\sqrt{2}}P_{+} \sim 71\% P_{+}$ tells us that the emitted radio power is much less dependent on $\rho_{S}$ for plasma densities larger than $\rho_{C}$. 
Furthermore, since $P_{-}$ decreases with decreasing $\rho_S$, then systems with higher plasma densities are more likely to be observable.

Regarding the orbital radius $r_{S}$, the only real physical constraints on the exomoon's orbit is that it must be close enough to the exoplanet to be well inside the magnetosphere, and gravitationally stable, but farther than the Roche limit to avoid structural instability.
However, we can also make use of Eq. (\ref{eq:P_rad_general}) to find orbits which might favor radio emissions, since exomoons in these orbits are the most likely to be detected.

\begin{figure}[PradvsrS]
\epsscale{0.80}
\includegraphics[width=0.85\linewidth]{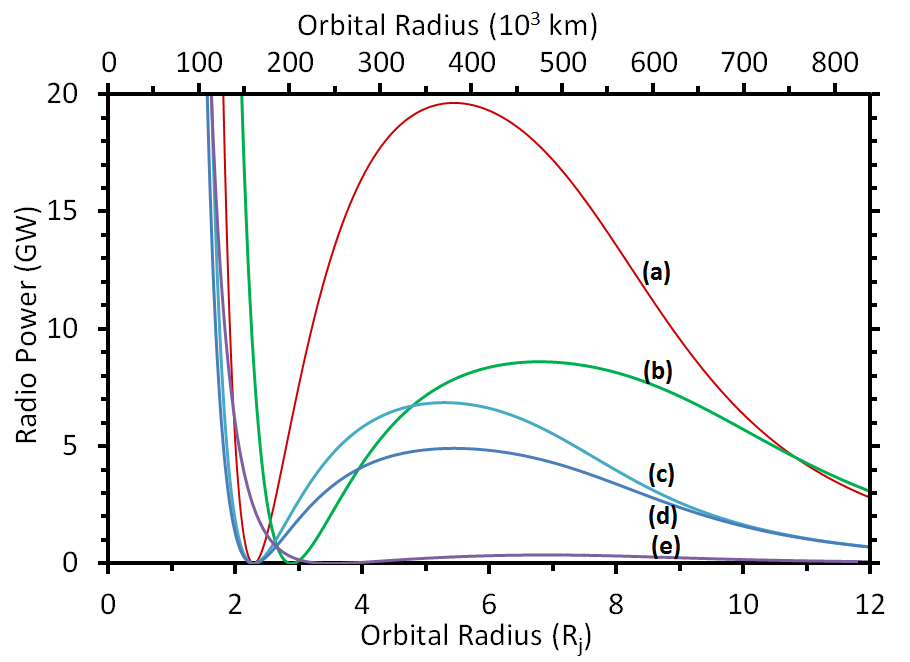}\hfill
\caption{$P_{rad}$ as a function of $r_{S}$, starting with values from the Jupiter-Io system, and changing one parameter at a time.
(a) $R_{S} = 2R_{Io}$, (b) $M_{P} = 2M_{J}$, (c) $\rho_{S} = 2\rho_{Io}$, (d) All Jupiter-Io values, (e) $T_{P} = 2T_{J}$.
The node seen at the synchronous orbit occurs because $V_{0}=0$ at this orbital distance.}
\label{fig:PradvsrS}
\end{figure}

The function $P_{rad}(r_{S})$ is plotted in Fig. \ref{fig:PradvsrS} for various parameter combinations.
The purpose of Fig. \ref{fig:PradvsrS} is to demonstrate how changing a single parameter in $P_{rad}$ affects its properties.
Clearly, $P_{rad}$ always has a single maximum in the orbits larger than the synchronous orbit. 
The only parameter that does not affect the orbital distance at which this maximum occurs is $R_{S}$; however, $\rho_{S}$ only affects the point's position weakly, and $T_{P}$ is always held constant, so $M_{P}$ is the dominant parameter when finding $P_{rad}$'s maximum.
The region around the maximum is also relatively flat; thereby, it seems reasonable to assume that an exomoon could be close to this maximum.  
In fact, Io and four Saturnian moons (including the large moon Enceladus) have orbits well within an area which would allow the moons to output at least 95\% 
of the maximum predicted radio power.
Therefore, we set $r_{S}$ to be the value that gives the maximum radio power, and rename it $r_{Opt}$ from this point forward to avoid confusion.  
We avoid the treatment of orbits smaller than the synchronous orbit due to their proximity to the Roche limit.

It is noteworthy that the optimal orbital radius for each planetary mass needs to be found numerically, but it approximately follows a power law. 
For example, using $\rho_{S} = \rho_{Io}$ we get $r_{Opt}=5.4M_{P}^{0.32}$, where $M_{P}$, and $r_{Opt}$ are given in units of Jupiter's mass, and radius.

\section{Detectability of exomoons}

For a power source a distance $d$ away from an observer, the incident flux is given by $S=P/(\Delta f \Omega d^{2})$, where $P$ is the source's output power, $\Delta f$ 
its bandwidth (usually taken to be half of the cyclotron frequency), and $\Omega$ is the solid angle through which the power is emitted by the source.
In the case of the Io-DAM, the emission cone half angle ranges from 60\degree to 90\degree, with a wall thickness of 1.5\degree\citep{lop07, que01}, which gives a solid angle of $\sim 0.14 - 0.16$ steradians.
Assuming a system also emitting with a wide half angle, and wall thickness up to 2$\degree$ gives $\Omega \sim 0.2$ steradians.
Taking $P_{rad}$ to be source's power, the incident flux becomes

\begin{equation}
S=\frac{2\pi\beta_{S}R_{S}^{2}B_{S}^{2}V_{0}}{\mu_{0}f_{C}\Omega d^{2}}\sqrt{\frac{\rho_{S}}{\rho_{S}+\mu_{0}^{-1}\left (B_{S}/V_{0}\right)^{2}}},
\label{eq:S1}
\end{equation}

\noindent
where $f_{C}$ is the cyclotron frequency of the system. 

\begin{figure}[fluxlines]
\epsscale{0.80}
\includegraphics[width=\textwidth]{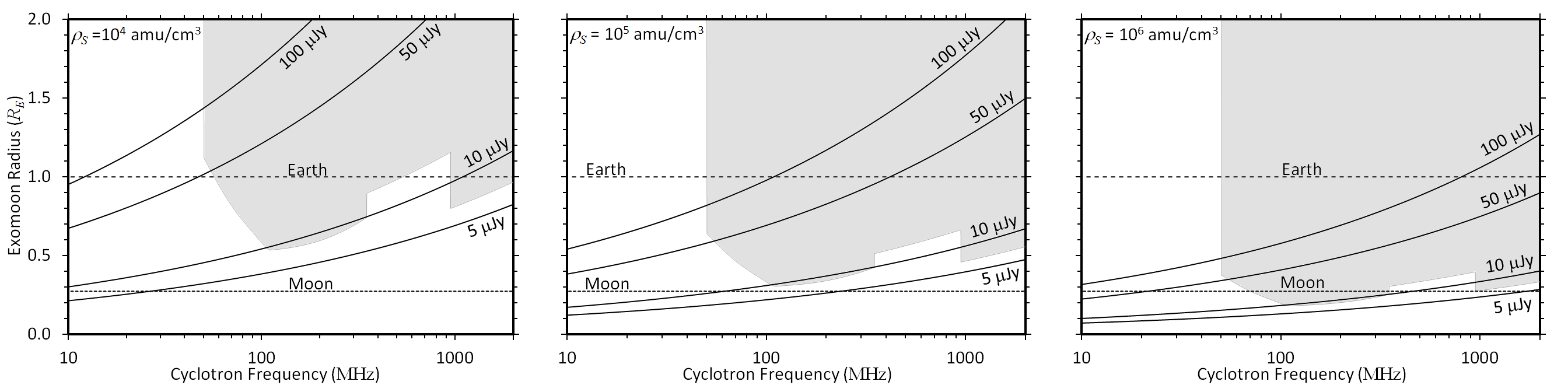}\hfill
\caption{Curves of output flux, $S$, plotted in the $R_{S}-f_{C}$ plane for several plasma densities.
Radii of detectable exomoons which are 15 light-years away plotted as a function of the host exoplanet's cyclotron frequency for several flux sensitivity values. 
From left to right, the panels show results for plasma densities of $10^{4}, 10^{5}, $ and $10^{6} amu/cm^{3}$.
The shaded area corresponds to the potential detection capabilities of a fully implemented SKA telescope.
1 Jansky = 10$^{-26}$ Wm$^{-2}$Hz.}
\label{fig:fluxlines}
\end{figure}

The cyclotron frequency is calculated using $f_{C} = eB_{pole}/(2\pi m_{e})$, where $e$ is the electron charge, $m_{e}$ is the electron mass, and we use the magnetic field stregth at the poles of the exoplanet, $B_{pole}$, because that is where most of the radio emissions occur.
At the poles of the planet, the magnetic field is twice as strong as it is at the equator.
Hence, we can use Eq. (\ref{eq:BS_final}) to express the strength of the magnetic field at the poles as $B_{pole}=2(r_{Opt}/R_{P})^{3}B_{S}$, where $R_{P}$ is the radius of the exoplanet.
Under this assumption, the cyclotron frequency increases as $M_{P}^{K}$, thereby limiting the amount of exoplanets that a telescope could successfully scan for exomoons. 
Nonetheless, there is still a wide range of frequencies within which an exomoon with radius $\le 1 R_{E}$ could be detected up to 15 light years away, if the plasma density $\rho_{S}$ of the system is at least $10^{4} amu$ $cm^{-3}$, and if the telescope's sensitivity is at least tens of $\mu$Jy (see Fig. \ref{fig:fluxlines}).
The proposed SKA telescope, if fully implemented, could even detect Mars-size moons ($\sim 0.532 R_{E}$)\footnote{Assuming 2 pol., 1 hr integration, and 16 MHz bandwidth} in this case, if present.
The range of detection of SKA is shown as the shaded areas shown in Fig. \ref{fig:fluxlines}.
Also, it must be noted that in reality these systems emit over a range of frequencies instead a single one, so the range of detectable systems is effectively larger than shown here.
The calculation of the whole frequency band will be treated in future studies.
Regarding $R_{P}$, a survey of currently known gas giants shows great variability in the value of this parameter. 
Nonetheless, for large planetary masses $R_{P}$ seems to converge to a value close to Jupiter's radius, $R_{J}$.
Furtermore, many authors (e.g. \cite{zar01}) assume that $R_{P} = R_{J}$ unless the exoplanet's radius is explicitly known.
Thus we will also assume $R_{P} = R_{J}$ to be the general case.

Applying our results to the exoplanet Epsilon Eridani b (1.55 $M_{J}$, 10.5 light years away), we find that a telescope with a flux sensitivity of $S\le$50 $\mu$Jy around 49 MHz could detect exomoons with radius between 0.24 $R_{E}$ for high $\rho_{S} (\sim 10^{6} amu$ $cm^{-3}$) and 0.73 $R_{E}$ for low $\rho_{S} (\sim 10^{4} amu$ $cm^{-3}$).
For comparison, The Moon is $\sim$0.273 $R_{E}$.
On another nearby exoplanet, Gliese 876 b (2.28 $M_{J}$, 15.29 light years away), a telescope with similar sensitivity around 93 MHz could detect an exomoon with a radius between $0.28$ and $0.86 R_{E}$, depending on $\rho_{S}$.
In both cases, a fairly large minimum radius is required for exomoons to be detectable, unless there is a large amount of plasma present.
In fact, Eq. (\ref{eq:S1}) tells us that to find an exomoon of radius 2500 km (similar to Mercury or Titan) orbiting Epsilon Eridani b, we would need a telescope with a flux sensitivity of 14 $\mu$Jy if $\rho_{S}$ is low.
Nevertheless, improvements to radio telescope technology, and observational techniques could one day make it possible to reach these sensitivities.

\section{Conclusions}

The primary goal of this study was to find a set of conditions that would allow detection of an exomoon through the radio emissions it induces on its host exoplanet, and to assess whether these conditions are attainable.
The results presented in Section 3 show that such conditions can exist, hence confirming the possibility of exomoon detection using this method, and we showed what sensitivities observational facilities need have to detect said exomoons.
An exomoon orbiting Epsilon Eridani b, with radius as low as 0.24 $R_{E}$, lies in the detectable range of telescopes with sensitivity $S\le$50 $\mu$Jy. 
However, detection of a Titan or Mercury-sized exomoon under low plasma conditions would require a telescope with flux sensitivity of $\sim 14$ $\mu$Jy or better. 

The model presented here still requires several refinements, such as including the effects of magnetic exomoons, finding better constraints on $T_{P}$ and $\rho_{S}$, and calculating a complete emission spectrum rather than a single cyclotron frequency. 
These improvements will be treated in later studies. 
Nonetheless, it is still our hope that the results presented here will give new insight to the observational community, and stimulate searches for the modulation of exo-planetary radio emissions caused by the presence of exomoons.

\emph{Acknowledgements}: We are grateful to the anonymous referee for his/her useful comments and suggestions which have allowed us to significantly improve this manuscript. We would also like to thank M. Cuntz for stimulating discussions and rendering his valuable comments. Our research on Alfv\'en waves in the exoplanet-exomoon environment was partially supported by NSF under the grant AGS 1246074 (ZEM and JPN), and GAANN Fellowship (JPN). ZEM also acknowledges the support of this work by the Alexander von Humboldt Foundation, and by University of Texas at Arlington through its Faculty Development Program.

{}

\clearpage

\end{document}